\documentclass[
superscriptaddress,
 amsmath,
 amssymb,
 aps,
 pra,
 twocolumn,
floatfix,
]{revtex4-2}

\usepackage{graphicx}
\usepackage{dcolumn}
\usepackage{bm}
\usepackage{physics}
\usepackage{natbib}
\usepackage{amsmath}

\begin{document}

\title{Coupling undetected sensing modes by quantum erasure}

\author{Nathan~R.~Gemmell}
\email{n.gemmell20@imperial.ac.uk}
\author{Yue~Ma}
\author{Emma~Pearce}%
\author{Jefferson~Fl\'orez}%
\author{Olaf~Czerwinski}
\author{M.~S.~Kim}
\author{Rupert~F.~Oulton}%
\affiliation{%
 Department of Physics, Blackett Laboratory, Imperial College London, South Kensington Campus, London SW7 2AZ, United Kingdom
}%

\author{Alex~S.~Clark}
\email{alex.clark@bristol.ac.uk} 
\affiliation{%
 Department of Physics, Blackett Laboratory, Imperial College London, South Kensington Campus, London SW7 2AZ, United Kingdom
}%
\affiliation{
 Quantum Engineering Technology Labs, H. H. Wills Physics Laboratory and Department of Electrical
and Electronic Engineering, University of Bristol, BS8 1FD, United Kingdom
}%
\author{Chris~C.~Phillips}%
\affiliation{%
 Department of Physics, Blackett Laboratory, Imperial College London, South Kensington Campus, London SW7 2AZ, United Kingdom
}%

\date{\today}

\begin{abstract} 
The effect known as ``induced coherence without induced emission'' has spawned a field dedicated to imaging with undetected photons (IUP), where photons from two distinct photon-pair sources interfere if their outputs are made indistinguishable. The indistinguishability is commonly achieved in two setups. Induced coherence IUP (IC-IUP) has only the idler photons from the first source passing through the second, whilst nonlinear interferometry (NI-IUP) has both signal and idler photons from the first source  passing through the second and can be simpler to implement. In both cases, changes in the idler path between sources can be detected by measuring the interference fringes in the signal path in a way that allows image information to be moved between different wavelengths. Here we model and implement a novel setup that uses a polarization state quantum eraser approach to move continuously between IC-IUP and NI-IUP operation. We find excellent agreement between experiment and theory in the low-gain or quantum regime. The  system also provides a new route for optimizing IUP interference by using controllable quantum erasure to balance the interferometer.

\end{abstract}
\maketitle

\section{Introduction}
In 1991, a paper by Zou, Wang, and Mandel demonstrated the capability of sensing changes in phase and transmission of a beam of light that would not itself ever be detected \cite{wang1991induced}. They achieved this remarkable feat by using two distinct photon-pair sources (nonlinear crystals in their example) which were pumped coherently using a single laser beam split into two. Each crystal produced a pair of outputs via spontaneous parametric down-conversion (SPDC) whose shorter/longer wavelengths were commonly known as the signal/idler beams. The idler output of the first crystal was directed through the second and the two signal beams were combined with a beamsplitter, making generation in the first or second crystal indistinguishable from one another, meaning it was impossible to determine which crystal a particular photon had come from. 
\begin{figure*}[t!]
\centering
\includegraphics[width=13cm]{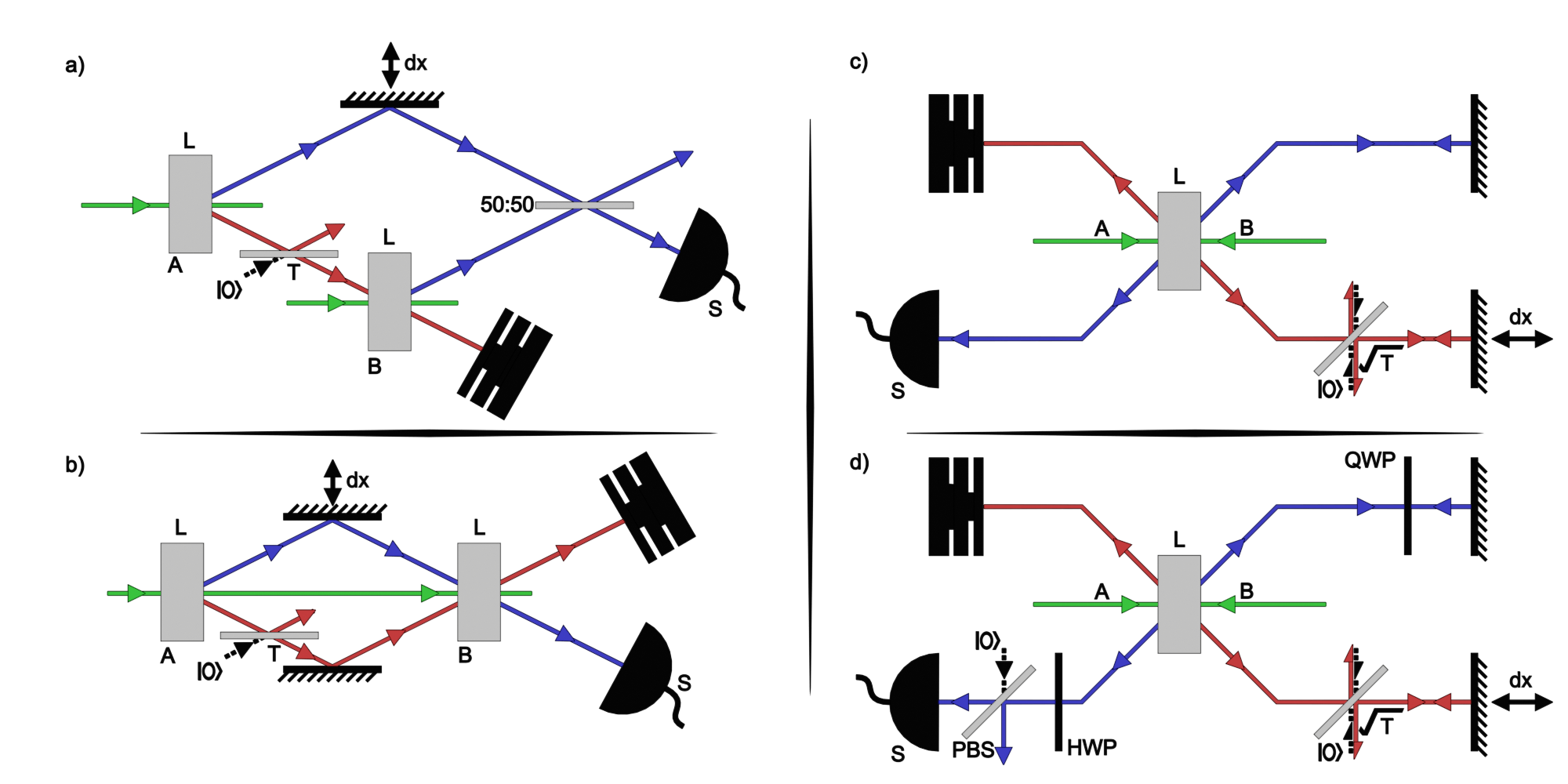}
\caption{Schematic of three imaging with undetected photons (IUP) theoretical models. (a) an induced coherence (IC-IUP) imaging setup, (b) a nonlinear interferometer (NI-IUP) setup, (c) a retro-reflected (RR-IUP)  setup, and (d) our model combining IC-IUP with NI-IUP in a single RR-IUP setup. Green, blue and red represent the pump, signal, and idler beams respectively. In all cases, process A is seeded only by vacuum, and detection takes place on the signal beam while the output idler photons are discarded. The crystal length is $L$, while the length of the idler arm is adjusted by length $dx$. The signal fields are mixed on a 50:50 beamsplitter in (a). In (d) the signal photons generated in process A pass twice through a quarter wave plate (QWP), and the signal fields after the output are mixed using a half waveplate (HWP) and a polarizing beamsplitter (PBS). In all cases a beamsplitter with transmission $T$ controls losses in the idler path.}
\label{fig:theory}
\end{figure*}
The resulting absence of ``which way'' information generates ``induced coherence'' and results in interference in all of the beams if one of the path lengths is scanned. An example of this setup is shown in \ref{fig:theory}(a). In this scenario, the interference fringes are recorded in the more readily-detected signal beam (usually the idler output is discarded). They have a phase that is dependent on the total phase between the two crystals, including that of the idler path, and an amplitude that is dependent on the degree of photon indistinguishability between the two crystal outputs. The latter is sensitive to how well the outputs of the two crystals have been overlapped optically, and it can be modulated by changing the transmission of the idler path between crystals \cite{Hochrainer2022}. 

In 2014, Lemos et al. demonstrated how to use this effect to create images. Combining the induced coherence effect  with the inherent anti-momentum correlations of the entangled photon pairs gave additional position information \cite{Lemos2014}. This ushered in a field of research, so-called ``Imaging with Undetected Photons'' (IUP). There are a variety of experimental configurations, but in each case the quantum correlations within the entangled photon pairs are harnessed to transfer image information from wavelengths where detectors may be expensive, slow, bulky, or noisy, to wavelengths that are much easier to detect. 

As there is some ambiguity in the terminology in the literature, here we define three particular configurations (\ref{fig:theory}). Figure~\ref{fig:theory}(a), IC-IUP, (where we use IC here to denote induced coherence), is the original imaging scheme pioneered by Lemos et al. Figure~\ref{fig:theory}(b) is a configuration often (but not exclusively) referred to as a nonlinear interferometer (NI-IUP), where both signal and idler photons from the first crystal are directed through the second \cite{Basset2021,Topfer2022,Kviatkovsky2020,Paterova2018,Lindner2020,Arahata2022,Rojas-Santana2022,Vanselow2020,Machado2020,Mukai2021,Cardoso2018,Michael2021,Vergyris2020,Kalashnikov2016,Lee2019,Buzas2020}. It should be noted though that the main reason this is done is to make it easier to overlap the beams from the two crystals accurately to achieve the photon indistinguishability that is required for interference. Usually these systems are operated in the low-gain (or quantum) regime, where only a single photon pair is in the system at any given time, and there is negligible interaction between the photon pairs generated in the first crystal with the other optical fields present in the second.   

Another experimental arrangement \cite{herzog1994} is shown in \ref{fig:theory}(c), which is now commonly used for IUP systems~\cite{Pearce2023}. In this retro-reflected (RR-IUP) scheme very stable photon indistinguishability can be achieved by using the same crystal twice, in an arrangement similar to a Michelson interferometer. All three beams -- pump, signal, and idler -- are reflected back through the crystal to close the interferometer. The idler path can be readily scanned, and still generates interference fringes in the signal beam, and this has resulted in some very compact and robust practical implementations \cite{Pearce2023}. 

While there have been individual attempts to characterize theoretically the IC-IUP and NI-IUP versions of this experiment \cite{Giese2017,Kolobov2017,Lemos2022}, to our knowledge the two have not been compared directly, either experimentally or theoretically.  Here we present a comparison of these two versions of IUP using a single experimental configuration by constructing a RR-IUP setup \cite{Gemmell2023} with a quantum erasure output (Fig.~\ref{fig:theory}(d)).  Quantum erasure is a term that describes the act of removing which-way (``welcher-weg'') distinguishing information from a system in a way that generates interference effects where none were previously present \cite{Zajonc1991}. In our setup, the signal beam from the first crystal pass traverses a quarter wave plate (QWP) twice, rotating its polarization in a way can introduce distinguishability between photons originating in the first and second passes of the crystal. A 0$^\circ$ rotation results in signal photons being aligned with those generated in the second crystal pass, creating a NI-IUP system. A 90$^\circ$ rotation gives perfect distinguishability between the two passes and stops the interference completely. Inserting a half wave plate (HWP) in the output signal beams can be used to rotate their polarizations by 45$^\circ$, in a way that means mixing can now occur between them at a polarizing beamsplitter, thus restoring the indistinguishability and therefore the interference \cite{herzog1995,kwiat1992,kwiat1994,Huo2022} forming an IC-IUP configuration. Here we also use this arrangement to investigate the effects of partial distinguishability and quantum erasure by continuously varying the QWP and HWP angles, thereby comparing the effect of IC-IUP and NI-IUP sensing modes and demonstrating phase-sensitive interference between the two interferometer configurations. The results agree well with our theoretical model and show that extra control in imaging and sensing systems can be achieved using this quantum erasure effect.

\section{Theory}

\begin{figure*}[t!]
\centering\includegraphics[width=14cm]{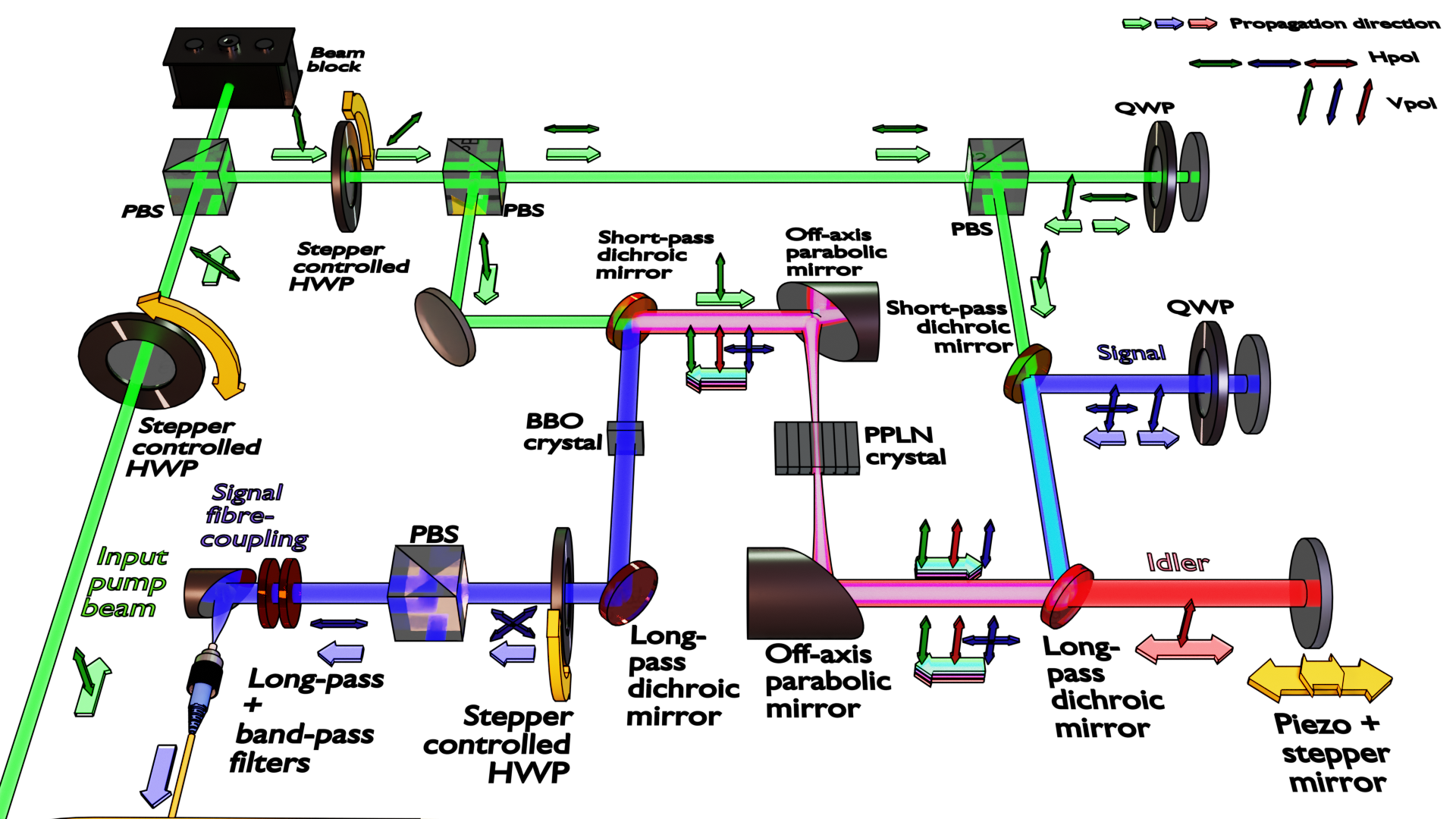}
\caption{Schematic of experimental setup. Green beam paths indicate pump (1064\,nm), blue represents the signal paths (1550\,nm), and red corresponds to the idler (3.4\,$\mu$m). Yellow arrows indicate automated translation.}
\label{fig:setup}
\end{figure*}

While both the IC-IUP and NI-IUP setups have been previously modelled \cite{Giese2017,Kolobov2017} the transition between the two modes of operation has not.  In the previous treatments simplifying assumptions (primarily single-mode and single frequency) have limited their ability to describe accurately the effects of introducing variable waveplates and polarizers.  

In both setups the path lengths of signal and idler photons between the crystals need to match to within the coherence length of the down-converted fields (typically $\sim0.1$~mm) in order that photon pairs from the first crystal pass cannot be distinguished from those from the second. From a theoretical point of view, this requirement is usually ignored, because in a single-frequency model the coherence length is infinite. However, in reality, finite frequency widths mean finite coherence lengths, and tight constraints on path lengths. These can be readily achieved in the standard setups,  but in our setup (Fig.~\ref{fig:theory}(d)) the birefringence of the nonlinear crystal must be taken into account as it causes nontrivial differences in the path lengths for photons that have had their polarizations rotated.  This difference needs to be compensated for by changing just one of the path lengths (signal or idler) between crystal passes, without changing the other. 

Our mathematical description of the system in Fig.~\ref{fig:theory}(d) is presented in full in Supplement 1. It includes the effects of finite frequency widths (and thus coherence lengths), path-length and phase shifts induced by crystal lengths, birefringence, and field polarization. The photon number output at detector S is described by:

\begin{widetext}
\begin{equation}
\begin{split}
N = \xi_{B}^{2}\cos^2{\theta_2}+\xi_{A}^2\cos^2{\theta_1}\cos^2{\theta_2}+\xi_{A}^2\sin^2{\theta_1}\sin^2{\theta_2}\\-2\xi_{A}^2\sin{\theta_1}\sin{\theta_2}\cos{\theta_1}\cos{\theta_2}\exp{-\frac{\Delta L_{H_s,V_s}^2}{2L_{coh}^2}}\cos{\{2\pi(\phi_{H_s}-\phi_{V_s})\}}\\+2\xi_{A}\xi_{B}\sqrt{T}\cos{\theta_1}\cos^2{\theta_2}\exp{-\frac{(\Delta L_{H_s,V,i}-dx)^2}{2L_{coh}^2}}\cos{\biggl\{2\pi\biggl(\phi_{H_s}+\phi_{V_i}+\frac{dx}{\lambda_i}\biggl)\biggl\}}\\-2\xi_{A}\xi_{B}\sqrt{T}\sin{\theta_1}\sin{\theta_2}\cos{\theta_2}\exp{-\frac{(\Delta L_{V_s,V,i}-dx)^2}{2L_{coh}^2}}\cos{\biggl\{2\pi\biggl(\phi_{V_s}+\phi_{V_i}+\frac{dx}{\lambda_i}\biggl)\biggl\}}
\end{split}
\label{eq:1}
\end{equation}
\end{widetext}

\noindent where $\xi_A$ and $\xi_B$ are gain parameters for first pass (A) and second pass (B) of the nonlinear crystal, $T$ is the transmissivity of the idler path between the two crystal passes, and $dx$ is the distance change of the idler mirror. The polarization rotation angles caused by the internal QWP and external HWP are represented by $\theta_{1}$ and $\theta_{2}$, respectively. The optical path length difference between polarization modes in the nonlinear crystal at the signal and idler wavelengths, $\lambda_s$ and $\lambda_i$ respectively, is defined as $\Delta L_{X_y,W_z} = L (n_{X_y}-n_{W_z})$ where $L$ is the crystal length and $n_{X_y}$ and $n_{W_z}$ are crystal refractive indices for polarization directions $H$ or $V$ (given as subscripts $X$ and $W$) at $\lambda_s$ or $\lambda_i$ (denoted by $s$ or $i$ for subscripts $y$ and $z$). The coherence length of the generated photons is defined as $L_{coh} = c/\Delta\omega_{s,i}$ where $c$ is the speed of light and $\Delta\omega_{s,i}$ is the photon frequency bandwidth. The accumulated phase between the exit of the crystal in the first (A) and second (B) passes is given by $\phi_{X_y}$ where once again $X\in\{H,V\}$ denotes the polarization and $y\in\{s,i\}$ the wavelength. Using Eq.\,\eqref{eq:1}, we can now model our experiment for arbitrary idler transmission, different internal QWP and external HWP angles to compare NI-IUP, IC-IUP, and combined setups, as well as uneven gain parameters for the two passes of the nonlinear crystal.

The strength of any interference effect can be defined using two parameters: the amplitude and the visibility of the interference fringes. The amplitude can be simply defined as the difference between the numbers of photons detected at a bright fringe ($N_\text{max}$) and those detected at a dark fringe ($N_\text{min}$) such that $\mathcal{A} = N_\text{max} - N_\text{min}$. The visibility of the interference fringes, $\mathcal{V}$, can then be defined simply as:

\begin{equation}
    \mathcal{V} = \frac{\mathcal{A}}{\left(N_\text{max} + N_\text{min}\right)}  \, .
\end{equation}

The visibility, $\mathcal{V}$, can be maximized by balancing the signal beam intensities that reach the detector from the two crystal passes.

\section{Experimental procedure}
Figure~\ref{fig:setup} shows a schematic of the optical setup designed to test our theory. A 1064 nm wavelength pump laser beam is set in power and polarization via an automated half-wave plate (HWP) and polarizing beamsplitter (PBS). A second HWP and PBS create the two pump beam paths for nonlinear processes A and B as described in the theory (the polarization of pump beam B is rotated to match A via a quarter-wave plate (QWP) and reflection from another PBS).

The two pump beams enter the crystal where signal and idler photons are generated through SPDC at wavelengths of 1550\,nm and 3.4\,$\mu$m respectively. The temperature-controlled periodically-poled lithium niobate (PPLN) crystal, supplied by Covesion, has a poling period of $\Lambda$ = 30.5\,$\mu$m, and anti-reflection coated to $R<1.5\%$ @ 1064\,nm, to $R<1\%$ @1400-1800\,nm, and to $R \sim 6\%-3\% $ @ 2600-4800\,nm, on both input/output facets. The pump beams are focused through two off-axis parabolic mirrors which also serve to collect and nominally collimate the signal and idler beams. Signal and idler photons from process A (the first pass) are separated from the other beams via a pair of dichroic mirrors and reflected back into the crystal along with the pump for process B (the second pass). 

A QWP in the signal arm allows for a polarization rotation to switch between the IC-IUP and NI-IUP operating modes. Signal and idler path lengths are matched via a computer-controlled stepper-motor translation stage, while phase shifts are implemented through a closed-loop piezoelectric translation stage, also in the idler path. The reflected signal beams from process A and from process B are separated from the pump with a third dichroic mirror. 

A 5\,mm long beta barium borate (BBO) crystal reduces the timing difference introduced between the orthogonal polarization components of the signal beam, whilst simultaneously allowing for the phase difference between the polarization components to be adjusted by tilting the crystal. An additional HWP and PBS before detection enables the two orthogonal components of the signal beam to be mixed. The signal is then filtered through a 2100\,nm longpass filter (Layertec), a 12 nm wide 1550\,nm bandpass filter (Thorlabs), and a 1400\,nm longpass filter (Thorlabs) before being coupled into a single-mode fibre (SMF-28). The signal beam is detected with a superconducting nanowire single-photon detector (SNSPD, IDQuantique, $\sim80$\% detection efficiency at 1550\,nm), with count rates monitored by a pulse counting module (ID900) with a 50\,ms acquisition time.

\begin{figure}[t!]
\centering\includegraphics[width=8cm]{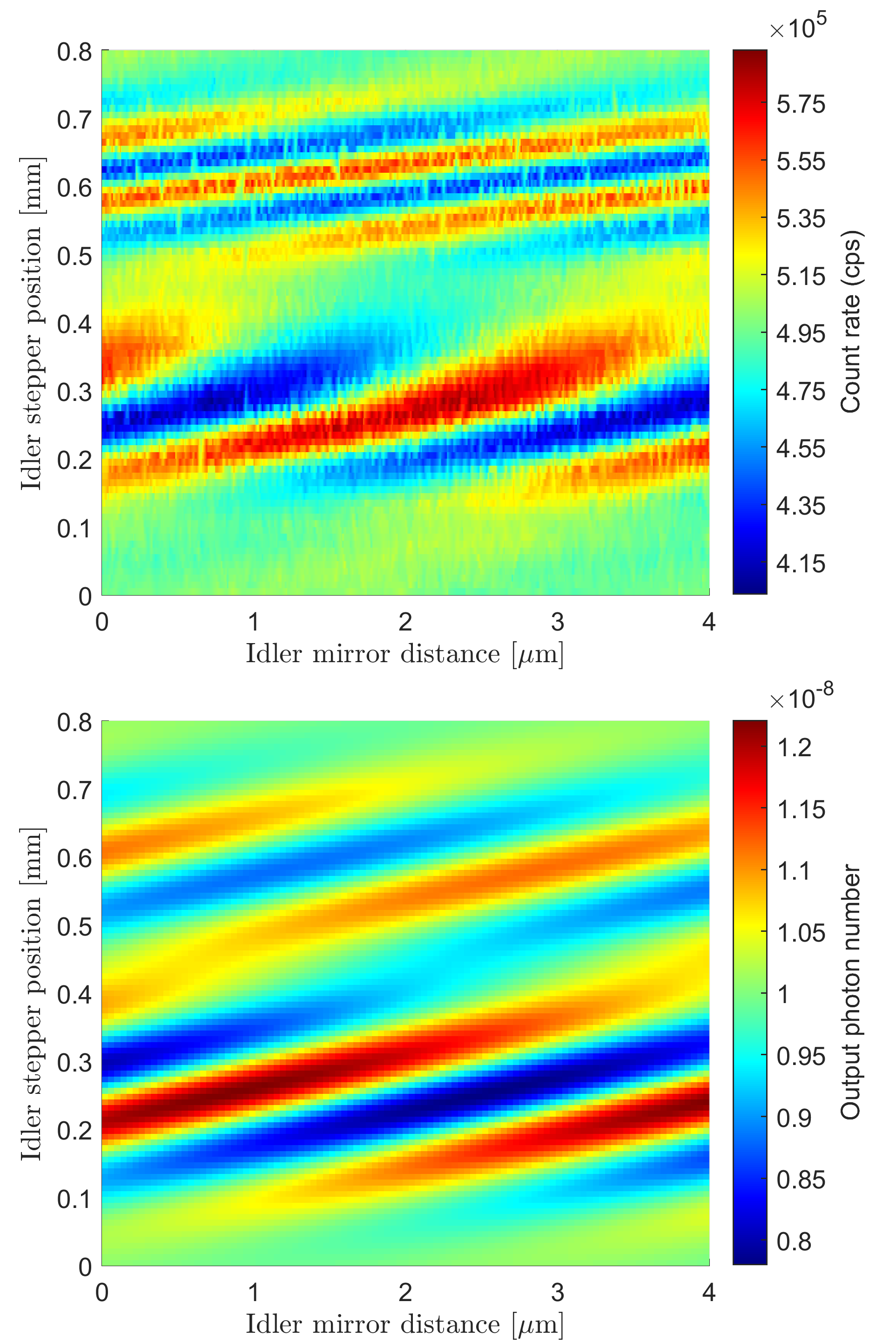}
\caption{Experimental signal count rates (top) and theoretical signal count rates (bottom) as a function of the large changes in idler path length on the y-axis and fine shifts in idler path length on the x-axis.}
\label{fig:exp2}
\end{figure}

\section{Results}

\begin{figure}[htbp!]
\centering\includegraphics[width=8cm]{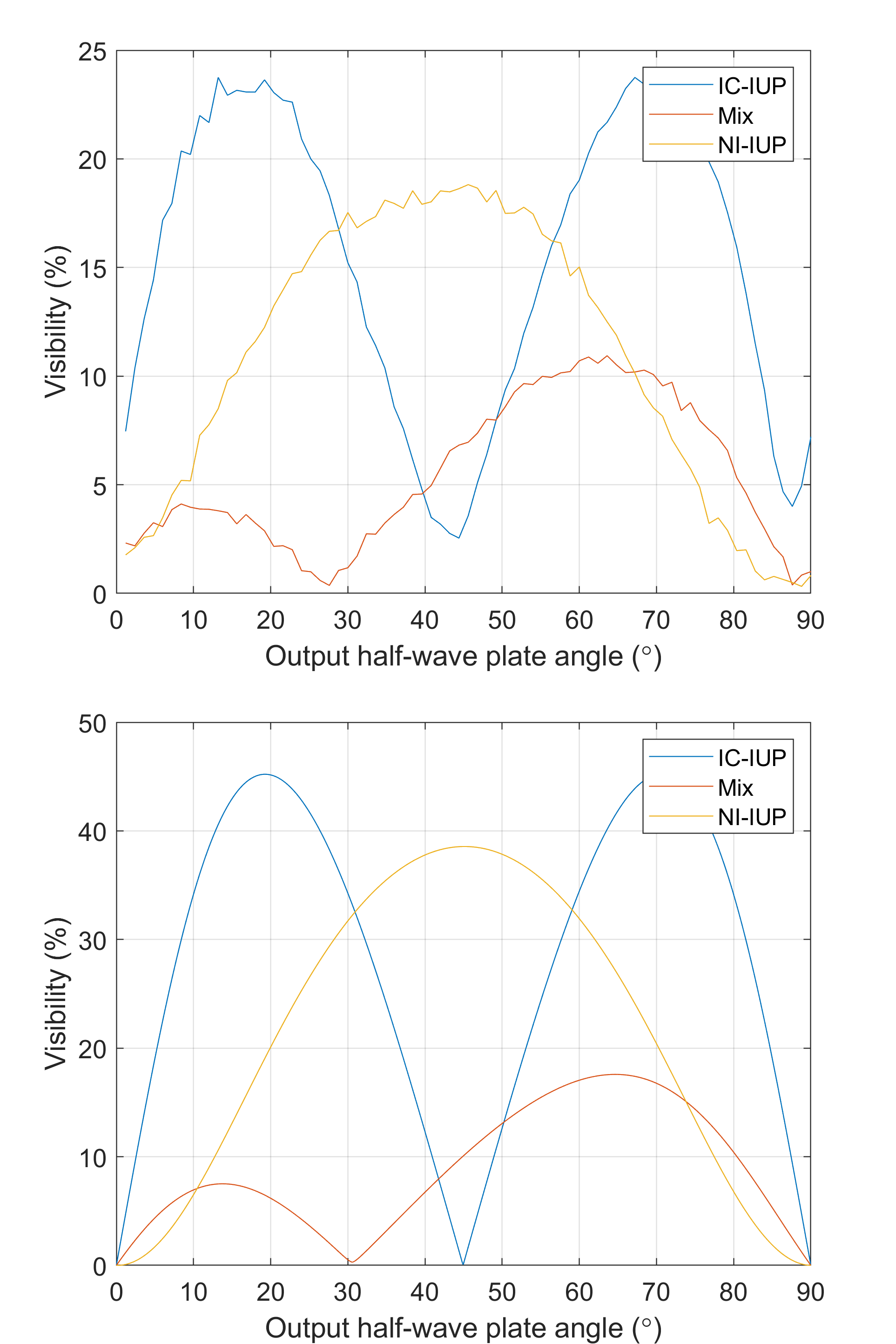}
\caption{Experimental data (top) and theoretical model (bottom) showing the visibility of the interference fringes as a function of the output HWP angle (corresponding to $\theta_2$). Data is shown for a single angle of the internal QWP and for three different idler path delays corresponding to IC-IUP in blue, NI-IUP in yellow, and a mixture in red. For the model, $T$ = 0.25, $\theta_1 = \sim30^{\circ}$.}
\label{fig:signalresults}
\end{figure}

The upper plot in Fig.~\ref{fig:exp2} shows experimental data, where the difference in length between the signal and idler paths has been changed by small (sub-wavelength)  shifts using the piezo-stage (plotted on the x-axis), as well as larger (mm) shifts using the stepper motor (plotted on the y-axis). Here the system has been put into hybrid state, part way between the IC-IUP and NI-IUP operating modes, by setting the internal QWP to $30^{\circ}$. This introduces partial distinguishability between photon pairs from the first and second passes of the crystal that is subsequently quantum erased by the external HWP and PBS. 

Two regions of interference, which occur at different macroscopic path length changes,  can be clearly identified; they correspond to the NI-IUP (at $\sim$0.25\,mm) and the IC-IUP (at $\sim$0.6\,mm) modes of operation. As discussed above, the change in refractive index experienced by the orthogonal component of the signal beam through the PPLN must be compensated for by changing the idler delay. The apparent changes in the ‘angle’ of the interference plots at the two macroscopic positions are artefacts caused by stepper motor non-linearities that cause aliasing effects in the data (see Supplement 1). The lower plot in Fig.~\ref{fig:exp2} shows the behavior of the systems as predicted by Eq.\ref{eq:1} for realistic input parameters ($T$ = 0.25, $L_{coh} = 0.2$\,mm). We note that the maximum observed visibility for the two cases is different, which we attribute to polarization dependent background levels in the experiment. In Supplement 1 we show background-free density plots of varying QWP and HWP angles for the IC-IUP and NI-IUP cases.

The upper plot in \ref{fig:signalresults} shows the visibilities of the interference fringes -- extracted from fits to experimental data -- at three different idler mirror delays which correspond to the NI-IUP, IC-IUP modes and part way in between. The lower plot shows the visibilities calculated from the corresponding theoretical model. These visibilities have been measured as a function of the angle of the output HWP ($\theta_2$), and they clearly demonstrate differences in how the interference is balanced in the NI-IUP and IC-IUP operating modes. Tuning the HWP can be used to balance the interference in the latter but not the former. 

In the NI-IUP case the only way to achieve a balance of counts from either crystal (and thereby maximise the fringe visibility) is to adjust the gain of one relative to the other, something that is difficult to do in the RR-IUP  setup. The IC-IUP mode offers the possibility of changing the beamsplitter ratio away from the 50:50 setting (as used by Lemos et al.); in our setup we can conveniently achieve the same effect simply by altering the HWP angle. 

Of additional interest is the region in between IC-IUP and NI-IUP modes, where the two modes of operation effectively interfere with each other. This data shows a fringe visibility minimum at a HWP angle of $\sim30^{\circ}$, and a maximum at $\sim65^{\circ}$ due to interference between the two modes of operation. The positions of the minimum and maximum visibility are dependent on the phase difference between the IC-IUP and the NI-IUP interference fringes. By tilting the BBO crystal it is possible to introduce a relative phase shift between the horizontal and vertical components of the signal photons, changing the interplay between the two interference effects, and adjusting the positions of the maximum and minimum (see Supplement 1).

\section{Conclusions}
We have shown,  both experimentally and theoretically, the differences between  IC-IUP and NI-IUP sensors in the low-gain/quantum regime. The two setups can be thought of as showing the same fundamental effect, although the IC-IUP setup offers the facility to rebalance the interference owing to the way the signal fields are combined using a beamsplitter. The IC-IUP mode also allows for the other exit port of the output beam splitter to be recorded with a second detector, potentially improving signal-to-noise ratios and data acquisition speeds. We have shown that, by introducing a controlled degree of quantum erasure of the distinguishing knowledge, it is possible to move between these modes and even sense or image in a hybrid mode. 

In the future we plan to investigate these effects in the high-gain regime, where we predict that the differences between the two modes will become more prominent, as the processes in the second crystal will become seeded by either one (IC-IUP) or both (NI-IUP) of the down-converted fields from the first crystal.

\section{Funding}

We acknowledge funding from the UK National Quantum Hub for Imaging (QUANTIC, No. EP/T00097X/1), an EPSRC DTP, and the Royal Society (No. UF160475).

\section{Data availability} Data underlying the results presented in this paper are not publicly available at this time but may be obtained from the authors upon reasonable request.

\bibliography{sample}
\clearpage
\pagebreak
\widetext
\begin{center}
\textbf{\large Supplemental Materials: Coupling undetected sensing modes by quantum erasure}
\end{center}
\setcounter{equation}{0}
\setcounter{figure}{0}
\setcounter{table}{0}
\setcounter{page}{1}
\setcounter{section}{0}
\makeatletter
\renewcommand{\theequation}{S\arabic{equation}}
\renewcommand{\thefigure}{S\arabic{figure}}
\renewcommand{\thesection}{S\arabic{section}}

\title{Supplemental content}
\maketitle
\section{Derivations}
We consider the experiment as described in Figure\,1(d) of the main manuscript. We will need to consider different frequencies for the two generated photons. However since spontaneous parametric downconversion (SPDC) is a pair-wise process, we can calculate for a single mode and sum over all modes at the end. We assume there is no polarization rotation of the idler field throughout the measurement, and so can consider only the horizontal component of the idler. Since our first process is unseeded, the initial input state is vacuum input into the two polarizations of the signal, the horizontally polarized idler, as well as vacuum into the second input of the idler mode beamsplitter which we later use for modelling loss in the idler arm of the interferometer: 
\begin{equation}
  \ket{0}_{H_s}\ket{0}_{V_s}\ket{0}_{H_i}\ket{0}_{b} 
\end{equation}
After the first SPDC process we generate the possibility of a photon pair into the horizontal signal and idler modes:
\begin{equation}
\begin{split}
\sim\ket{0}_{H_s}\ket{0}_{V_s}\ket{0}_{H_i}\ket{0}_{b}\\ + \xi_A\ket{1}_{H_s}\ket{0}_{V_s}\ket{1}_{H_i}\ket{0}_{b} 
\end{split}
\end{equation}
Since our SPDC process is type 0, signal and idler field polarizations match that of the pump. After including the path length change ($d_x$) introduced on the idler (of frequency $\omega_i$) path:
\begin{equation}
\begin{split}
\sim\ket{0}_{H_s}\ket{0}_{V_s}\ket{0}_{H_i}\ket{0}_{b}\\ + \xi_A\exp{-i\omega_i\frac{d_x}{c}}\ket{1}_{H_s}\ket{0}_{V_s}\ket{1}_{H_i}\ket{0}_{b} 
\end{split}
\end{equation}
A transmissive object modelled as a beamsplitter in the idler mode has transmissivity $T = \cos^2(\beta)$. After passing the beamsplitter:
\begin{equation}
\begin{split}
\sim\ket{0}_{H_s}\ket{0}_{V_s}\ket{0}_{H_i}\ket{0}_{b}\\ + \xi_A\exp{-i\omega_i\frac{d_x}{c}}\cos{\beta}\ket{1}_{H_s}\ket{0}_{V_s}\ket{1}_{H_i}\ket{0}_{b}\\ + \xi_A\exp{-i\omega_i\frac{d_x}{c}}\sin{\beta}\ket{1}_{H_s}\ket{0}_{V_s}\ket{0}_{H_i}\ket{1}_{b} 
\end{split}
\end{equation}

\noindent The double-passed quarter wave plate acts as a single pass half wave plate in the signal mode, with rotation angle $\theta_1$, enabling a change between $H_s$ and $V_s$:

\begin{equation}
\begin{split}
\sim\ket{0}_{H_s}\ket{0}_{V_s}\ket{0}_{H_i}\ket{0}_{b} + \xi_A\exp{-i\omega_i\frac{d_x}{c}}\cos{\beta}\cdot \cos{\theta_1}\ket{1}_{H_s}\ket{0}_{V_s}\ket{1}_{H_i}\ket{0}_{b} \\
+ \xi_A\exp{-i\omega_i\frac{d_x}{c}}\cos{\beta}\cdot i\sin{\theta_1}\ket{0}_{H_s}\ket{1}_{V_s}\ket{1}_{H_i}\ket{0}_{b}\\
+ \xi_A\exp{-i\omega_i\frac{d_x}{c}}\sin{\beta}\cdot \cos{\theta_1}\ket{1}_{H_s}\ket{0}_{V_s}\ket{0}_{H_i}\ket{1}_{b}\\
+ \xi_A\exp{-i\omega_i\frac{d_x}{c}}\sin{\beta}\cdot i\sin{\theta_1}\ket{0}_{H_s}\ket{1}_{V_s}\ket{0}_{H_i}\ket{1}_{b}
\end{split}
\end{equation}

\noindent Signal and idler modes are then remixed in the second SPDC process generating $\ket{1}_{H_s}\ket{1}_{H_i}$. Due to the birefringence of the second crystal (of length $L$), horizontal and vertical components of signal modes will experience difference delays due to refractive index changes. Adding in the additional generation and phase shifts gives:

\begin{equation}
\begin{split}
\sim\ket{0}_{H_s}\ket{0}_{V_s}\ket{0}_{H_i}\ket{0}_{b}\\
+ \xi_B\ket{1}_{H_s}\ket{0}_{V_s}\ket{1}_{H_i}\ket{0}_{b}\\
+ \xi_A\exp{-i\biggl(\omega_i\frac{d_x}{c} + \omega_s\frac{L}{c_{H_s}} + \omega_i\frac{L}{c_{H_i}}\biggl)}\cos{\beta}\cdot \cos{\theta_1}\ket{1}_{H_s}\ket{0}_{V_s}\ket{1}_{H_i}\ket{0}_{b} \\
+ \xi_A\exp{-i\biggl(\omega_i\frac{d_x}{c}+ + \omega_s\frac{L}{c_{V_s}} + \omega_i\frac{L}{c_{H_i}}\biggl)}\cos{\beta}\cdot i\sin{\theta_1}\ket{0}_{H_s}\ket{1}_{V_s}\ket{1}_{H_i}\ket{0}_{b}\\
+ \xi_A\exp{-i\biggl(\omega_i\frac{d_x}{c} + \omega_s\frac{L}{c_{H_s}}\biggl)}\sin{\beta}\cdot \cos{\theta_1}\ket{1}_{H_s}\ket{0}_{V_s}\ket{0}_{H_i}\ket{1}_{b}\\
+ \xi_A\exp{-i\biggl(\omega_i\frac{d_x}{c} + \omega_s\frac{L}{c_{V_s}}\biggl)}\sin{\beta}\cdot i\sin{\theta_1}\ket{0}_{H_s}\ket{1}_{V_s}\ket{0}_{H_i}\ket{1}_{b}
\end{split}
\end{equation}

\noindent and after the second HWP (with angle of rotation of $\theta_2$), the state can be found to be:

\begin{equation}
\begin{split}
\sim\ket{0}_{H_s}\ket{0}_{V_s}\ket{0}_{H_i}\ket{0}_{b}\\ 
+ (\xi_B\cos{\theta_2}+\xi_A\exp{-i\biggl(\omega_i\frac{d_x}{c} + \omega_s\frac{L}{c_{H_s}} + \omega_i\frac{L}{c_{H_i}}\biggl)}\cos{\beta}\cdot \cos{\theta_1}\cdot \cos{\theta_2}\\
+ \xi_A\exp{-i\biggl(\omega_i\frac{d_x}{c} + \omega_s\frac{L}{c_{V_s}} + \omega_i\frac{L}{c_{H_i}}\biggl)}\cos{\beta}\cdot i\sin{\theta_1}\cdot i\sin{\theta_2})\ket{1}_{H_s}\ket{0}_{V_s}\ket{1}_{H_i}\ket{0}_{b} \\
+ (\xi_Bi\sin{\theta_2} + \xi_A\exp{-i\biggl(\omega_i\frac{d_x}{c} + \omega_s\frac{L}{c_{H_s}} + \omega_i\frac{L}{c_{H_i}}\biggl)}\cos{\beta}\cdot \cos{\theta_1}\cdot i\sin{\theta_2}\\
+\xi_A\exp{-i\biggl(\omega_i\frac{d_x}{c} + \omega_s\frac{L}{c_{V_s}} + \omega_i\frac{L}{c_{H_i}}\biggl)}\cos{\beta}\cdot i\sin{\theta_1}\cdot \cos{\theta_2})\ket{0}_{H_s}\ket{1}_{V_s}\ket{1}_{H_i}\ket{0}_{b}\\
+ (\xi_A\exp{-i\biggl(\omega_i\frac{d_x}{c} + \omega_s\frac{L}{c_{H_s}}\biggl)}\sin{\beta}\cdot \cos{\theta_1}\cdot \cos{\theta_2}\\ + \xi_A\exp{-i\biggl(\omega_i\frac{d_x}{c} + \omega_s\frac{L}{c_{V_s}}\biggl)}\sin{\beta}\cdot i\sin{\theta_1}\cdot i\sin{\theta_2})\ket{1}_{H_s}\ket{0}_{V_s}\ket{0}_{H_i}\ket{1}_{b}\\
+ (\xi_A\exp{-i\biggl(\omega_i\frac{d_x}{c} + \omega_s\frac{L}{c_{H_s}}\biggl)}\sin{\beta}\cdot \cos{\theta_1}\cdot i\sin{\theta_2}\\
+ \xi_A\exp{-i\biggl(\omega_i\frac{d_x}{c} + \omega_s\frac{L}{c_{V_s}}\biggl)}\sin{\beta}\cdot i\sin{\theta_1}\cdot \cos{\theta_2})\ket{0}_{H_s}\ket{1}_{V_s}\ket{0}_{H_i}\ket{1}_{b}\\
\end{split}
\end{equation}

\noindent After the final PBS, we can now measure the H-polarized signal photon number state at detector S:

\begin{equation}
\begin{split}
\langle a_{H_s}^\dagger a_{H_s} \rangle = \xi_B^2\cos^2{\theta_2}+\xi_A^2\cos^2{\theta_1}\cos^2{2\theta_2} + \xi_A^2\sin^2{\theta_1}\sin^2{\theta_2}\\ 
+ \xi_A^2\sin^2{\beta}\sin^2{\theta_1}\sin^2{\theta_2}\\ 
+ 2\xi_A\xi_B\cos{\biggl\{\omega_i\frac{d_x}{c}+\omega_s\frac{L}{c_{H_s}}+\omega_i\frac{L}{c_{H_i}}\biggl\}}\cos{\beta}\cos{\theta_1}\cos^2{\theta_2}\\
- 2\xi_A\xi_B\cos{\biggl\{\omega_i\frac{d_x}{c}+\omega_s\frac{L}{c_{V_s}}+\omega_i\frac{L}{c_{H_i}}\biggl\}}\cos{\beta}\sin{\theta_1}\sin{\theta_2}\cos{\theta_2}\\
- 2\xi_A^2\cos{\biggl\{\omega_s\frac{L}{c_{H_s}}-\omega_s\frac{L}{c_{V_s}}\biggl\}}\sin{\theta_1}\cos{\theta_1}\sin{\theta_2}\cos{\theta_2}
\end{split}
\end{equation}

\noindent We assume that the gain of each of the nonlinear processes $\xi_j$ at signal frequency $\omega_s$ can be expressed using $\xi_{j0}$, the gain of the phase-matched signal frequency $w_{s0}$, and the coherence width of the down-converted signal, $\Delta\omega_s$ (assuming a Gaussian coherence envelope):

\begin{equation}
    \xi_j^2 = \xi_{j0}^2 \cdot \frac{1}{\sqrt{2\pi}\Delta\omega_s}\cdot \exp{-\frac{(\omega_s-\omega_{s0})^2}{2\Delta\omega_s^2}}
\end{equation}
\noindent where $j \in \{A,B\}$. From here, we can now make use of the pairwise property ($\omega_i = \omega_p - \omega_s$, where $\omega_x$) of SPDC and integrate over all frequencies.

\begin{equation}
\begin{split}
\int d\omega_s\braket{a_{H_s}^\dagger(\omega_s) a_{H_s}(\omega_s)} = \xi_{B0}^{2}\cos^2{\theta_2}+\xi_{A0}^2\cos^2{\beta}\cos^2{\theta_1}\cos^2{\theta_2}\\+\xi_{A0}^2\cos^2{\beta}\sin^2{\theta_1}\sin^2{\theta_2}+\xi_{A0}^2\sin^2{\beta}\cos^2{\theta_1}\cos^2{\theta_2}+\xi_{A0}^2\sin^2{\beta}\sin^2{\theta_1}\sin^2{\theta_2}\\+\int2\xi_A\xi_B\cos{\biggl\{\omega_p\biggl(\frac{d_x}{c}+\frac{L}{c_{H_i}}\biggl)+\omega_s\biggl(\frac{L}{c_{H_s}}-\frac{d_x}{c}-\frac{L}{c_{H_i}}\biggl)\biggl\}}\cos{\beta}\cos{\theta_1}\cos^2{\theta_2}d\omega\\+\int2\xi_A\xi_B\cos{\biggl\{\omega_p\biggl(\frac{d_x}{c}+\frac{L}{c_{H_i}}\biggl)+\omega_s\biggl(\frac{L}{c_{V_s}}-\frac{d_x}{c}-\frac{L}{c_{H_i}}\biggl)\biggl\}}\cos{\beta}\sin{\theta_1}\sin{\theta_2}\cos{\theta_2}d\omega\\+\int2\xi_A^2\cos{\biggl\{\omega_s\biggl(\frac{L}{c_{H_s}}-\frac{L}{c_{V_s}}\biggl)\biggl\}}\cos^2{\beta}\sin{\theta_1}\cos{\theta_1}\sin{\theta_2}\cos{\theta_2}d\omega\\+\int2\xi_A^2\cos{\biggl\{\omega_s\biggl(\frac{L}{c_{H_s}}-\frac{L}{c_{V_s}}\biggl)\biggl\}}\sin^2{\beta}\sin{\theta_1}\cos{\theta_1}\sin{\theta_2}\cos{\theta_2}d\omega\\
\end{split}
\end{equation}

\noindent Making use of the integral:

\begin{equation}
\int_{-\infty}^{+\infty}dx\frac{1}{\sqrt{2\pi}\Delta}\exp{-\frac{x^2}{2\Delta^2}}\cos{(y+xa)} = \exp{-\frac{1}{2}a^2\Delta^2}\cos{y}
\end{equation}

\noindent we find that the expectation value of the output at detector S can be given by:

\begin{equation}
\begin{split}
\braket{N} = \xi_{B0}^{2}\cos^2{\theta_2}+\xi_{A0}^2\cos^2{\theta_1}\cos^2{\theta_2}+\xi_{A0}^2\sin^2{\theta_1}\sin^2{\theta_2}\\-2\xi_{A0}^2\sin{\theta_1}\sin{\theta_2}\cos{\theta_1}\cos{\theta_2}\exp{-\frac{1}{2}\Delta\omega_s^2\biggl(\frac{L}{c_{H_s}}-\frac{L}{c_{V_s}}\biggl)^2}\cos{\biggl\{\omega_{s0}\biggl(\frac{L}{c_{H_s}}-\frac{L}{c_{V_S}}\biggl)\biggl\}}\\
+2\xi_{A0}\xi_{B0}\cos{\beta}\cos{\theta_1}\cos^2{\theta_2}\exp{-\frac{1}{2}\Delta\omega_s^2\biggl(\frac{L}{c_{H_s}}-\frac{dx}{c}-\frac{L}{c_{H_i}}\biggl)^2}\cos{\biggl\{\omega_{s0}\frac{L}{c_{H_s}}+(\omega_p-\omega_{s0})\biggl(\frac{d_x}{c}-\frac{L}{c_{H_i}}\biggl)\biggl\}}\\
-2\xi_{A0}\xi_{B0}\cos{\beta}\sin{\theta_1}\sin{\theta_2}\cos{\theta_2}\exp{-\frac{1}{2}\Delta\omega_1^2\biggl(\frac{L}{c_{V_s}}-\frac{d_x}{c}+\frac{L}{c_{H_i}}\biggl)^2}\cos{\biggl\{\omega_{s0}\frac{L}{c_{V_s}}+(\omega_p-\omega_{s0})\biggl(\frac{d_x}{c}+\frac{L}{c_{H_i}}\biggl)\biggl\}}
\end{split}
\end{equation}

\noindent Remembering that $\omega_{i0} = \omega_p-\omega_{s0}$, and recognizing that $c_{X_y} = c/n_{X,y}$ where $n_{X_y}$ are crystal refractive indices for polarization directions $H$ or $V$ given as subscript $X$ at $\lambda_s$ or $\lambda_i$ denoted by $s$ or $i$ for subscripts $y$, we can rewrite:

\begin{equation}
\begin{split}
\braket{N} = \xi_{B}^{2}\cos^2{\theta_2}+\xi_{A}^2\cos^2{\theta_1}\cos^2{\theta_2}+\xi_{A}^2\sin^2{\theta_1}\sin^2{\theta_2}\\
-2\xi_{A}^2\sin{\theta_1}\sin{\theta_2}\cos{\theta_1}\cos{\theta_2}\exp{-\frac{\Delta\omega_s^2L^2}{2c^2}(n_{H_s}-n_{V_s})^2}\cos{\biggl\{\frac{\omega_{s}L}{c}(n_{H_s}-n_{V_s})\biggl\}}\\
+2\xi_{A}\xi_{B}\cos{\beta}\cos{\theta_1}\cos^2{\theta_2}\exp{-\frac{\Delta\omega_1^2}{2c^2}(Ln_{H_s}-Ln_{V_i}-d_x)^2}\cos{\biggl\{\frac{\omega_{s}Ln_{H_s}+\omega_iLn_{V_i}+\omega_id_x}{c}\biggl\}}\\
-2\xi_{A}\xi_{B}\cos{\beta}\sin{\theta_1}\sin{\theta_2}\cos{\theta_2}\exp{-\frac{\Delta\omega_1^2}{2c^2}(Ln_{v_s}-Ln_{V_i}-d_x)^2}\cos{\biggl\{\frac{\omega_{s}Ln_{v_s}+\omega_iLn_{V_i}+\omega_id_x}{c}\biggl\}}
\end{split}
\end{equation}

\noindent We identify the coherence length of the generated photons as $L_{coh} = c/\Delta\omega_{s,i}$, and define the optical path length difference between polarization modes in the nonlinear crystal at the signal and idler wavelengths, $\lambda_s$ and $\lambda_i$ respectively, is defined as $\Delta L_{X_y,W_z} = L (n_{X_y}-n_{W_z})$. We also replace angular frequencies with wavelengths to give:

\begin{equation}
\begin{split}
\braket{N} = \xi_{B}^{2}\cos^2{\theta_2}+\xi_{A}^2\cos^2{\theta_1}\cos^2{\theta_2}+\xi_{A}^2\sin^2{\theta_1}\sin^2{\theta_2}\\-2\xi_{A}^2\sin{\theta_1}\sin{\theta_2}\cos{\theta_1}\cos{\theta_2}\exp{-\frac{\Delta L_{H_s,V_s}^2}{2L_{coh}^2}}\cos{\biggl\{\frac{2\pi \Delta L_{H_s,V,s}}{\lambda_s}\biggl\}}\\+2\xi_{A}\xi_{B}\cos{\beta}\cos{\theta_1}\cos^2{\theta_2}\exp{-\frac{(\Delta L_{H_s,V,i}-dx)^2}{2L_{coh}^2}}\cos{\biggl\{2\pi\biggl(\frac{Ln_{H_s}}{\lambda_s}+\frac{Ln_{V_i}}{\lambda_i}+\frac{dx}{\lambda_i}\biggl)\biggl\}}\\-2\xi_{A}\xi_{B}\cos{\beta}\sin{\theta_1}\sin{\theta_2}\cos{\theta_2}\exp{-\frac{(\Delta L_{V_s,V,i}-dx)^2}{2L_{coh}^2}}\cos{\biggl\{2\pi\biggl(\frac{Ln_{v_s}}{\lambda_s}+\frac{Ln_{V_i}}{\lambda_i}+\frac{dx}{\lambda_i}\biggl)\biggl\}}
\end{split}
\end{equation}

\noindent Finally we define the accumulated phase between the exit of the crystal in the first (A) and second (B) passes as $\phi_{X_y} = Ln_{X_y}/\lambda_y$ where once again $X\in\{H,V\}$ denotes the polarization and $y\in\{s,i\}$ the wavelength.

\begin{equation}
\begin{split}
\braket{N} = \xi_{B}^{2}\cos^2{\theta_2}+\xi_{A}^2\cos^2{\theta_1}\cos^2{\theta_2}+\xi_{A}^2\sin^2{\theta_1}\sin^2{\theta_2}\\-2\xi_{A}^2\sin{\theta_1}\sin{\theta_2}\cos{\theta_1}\cos{\theta_2}\exp{-\frac{\Delta L_{H_s,V_s}^2}{2L_{coh}^2}}\cos{\{2\pi(\phi_{H_s}-\phi_{V_s})\}}\\+2\xi_{A}\xi_{B}\sqrt{T}\cos{\theta_1}\cos^2{\theta_2}\exp{-\frac{(\Delta L_{H_s,V,i}-dx)^2}{2L_{coh}^2}}\cos{\biggl\{2\pi\biggl(\phi_{H_s}-\phi_{V_i}+\frac{dx}{\lambda_i}\biggl)\biggl\}}\\-2\xi_{A}\xi_{B}\sqrt{T}\sin{\theta_1}\sin{\theta_2}\cos{\theta_2}\exp{-\frac{(\Delta L_{V_s,V,i}-dx)^2}{2L_{coh}^2}}\cos{\biggl\{2\pi\biggl(\phi_{V_s}-\phi_{V_i}+\frac{dx}{\lambda_i}\biggl)\biggl\}}
\end{split}
\end{equation}


\section{Balancing}

As shown in \cite{Gemmell2023} it can be desirable to purposefully unbalance the undetected imaging setup in order to reduce the the total optical power that probes a sample. This was shown to come at the expense of visibility. With the additional balancing pathway offered by the induced coherence setup and PBS mixing however, this loss of visibility can be mitigated.

\begin{figure}[h!]
\centering\includegraphics[width = 16cm]{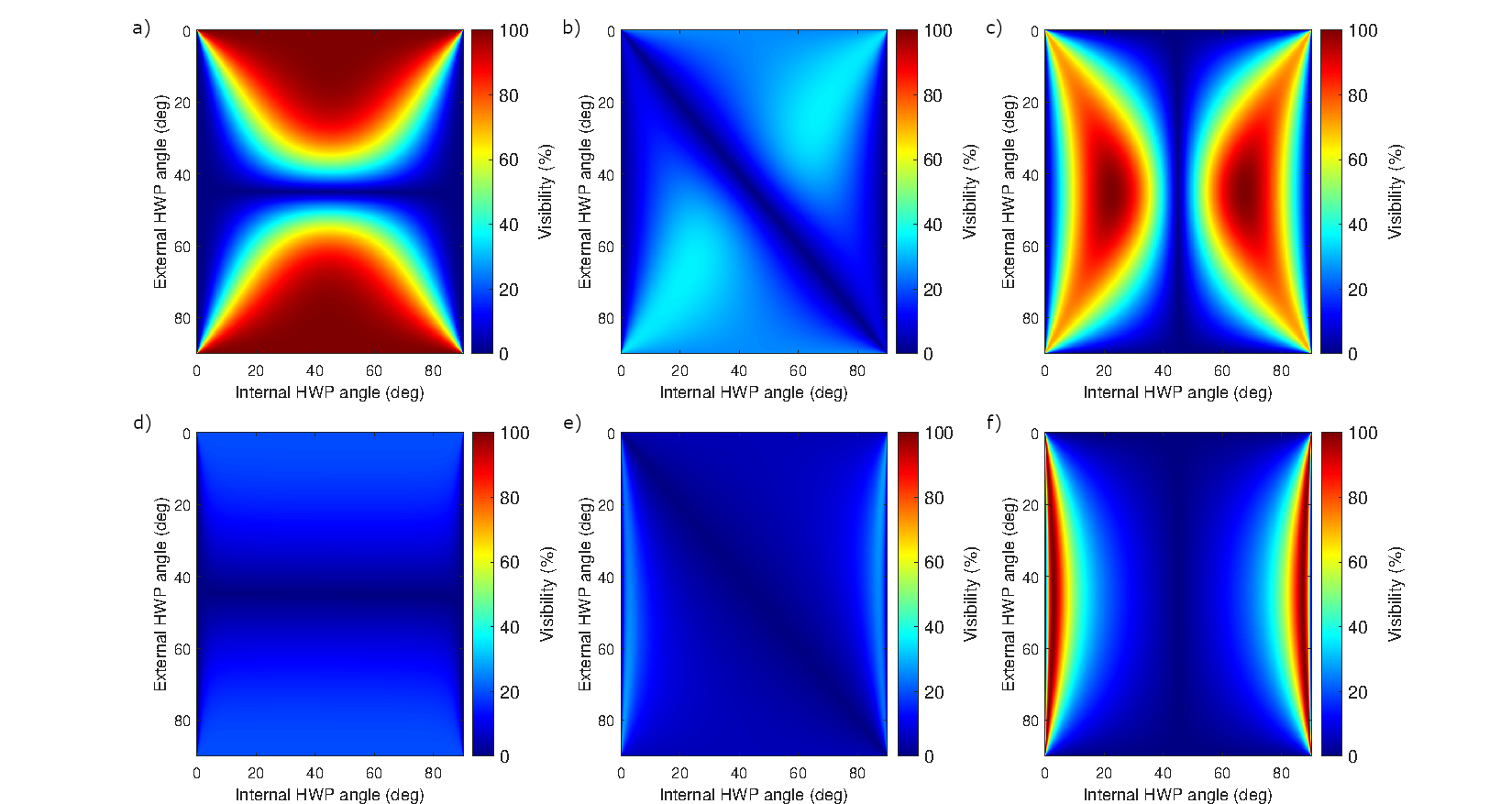}
\caption{Density plots of the visibility as a function of both of the HWP angles within the setup for three different idler mirror positions and two different gain balances. Upper plots show the visibilities for (a) nonlinear interference (NI-IUP), (b) the mixed case, and (c) induced coherence (IC-IUP) respectively for $\xi_A == \xi_B$. Lower plots show the visibilities for (d) NI-IUP, (e) the mixed case, and (f) IC-IUP respectively for $\xi_A = 0.1\xi_B$}
\label{fig:Balancing1}
\end{figure}

\begin{figure}[h!]
\centering\includegraphics[width = 16cm]{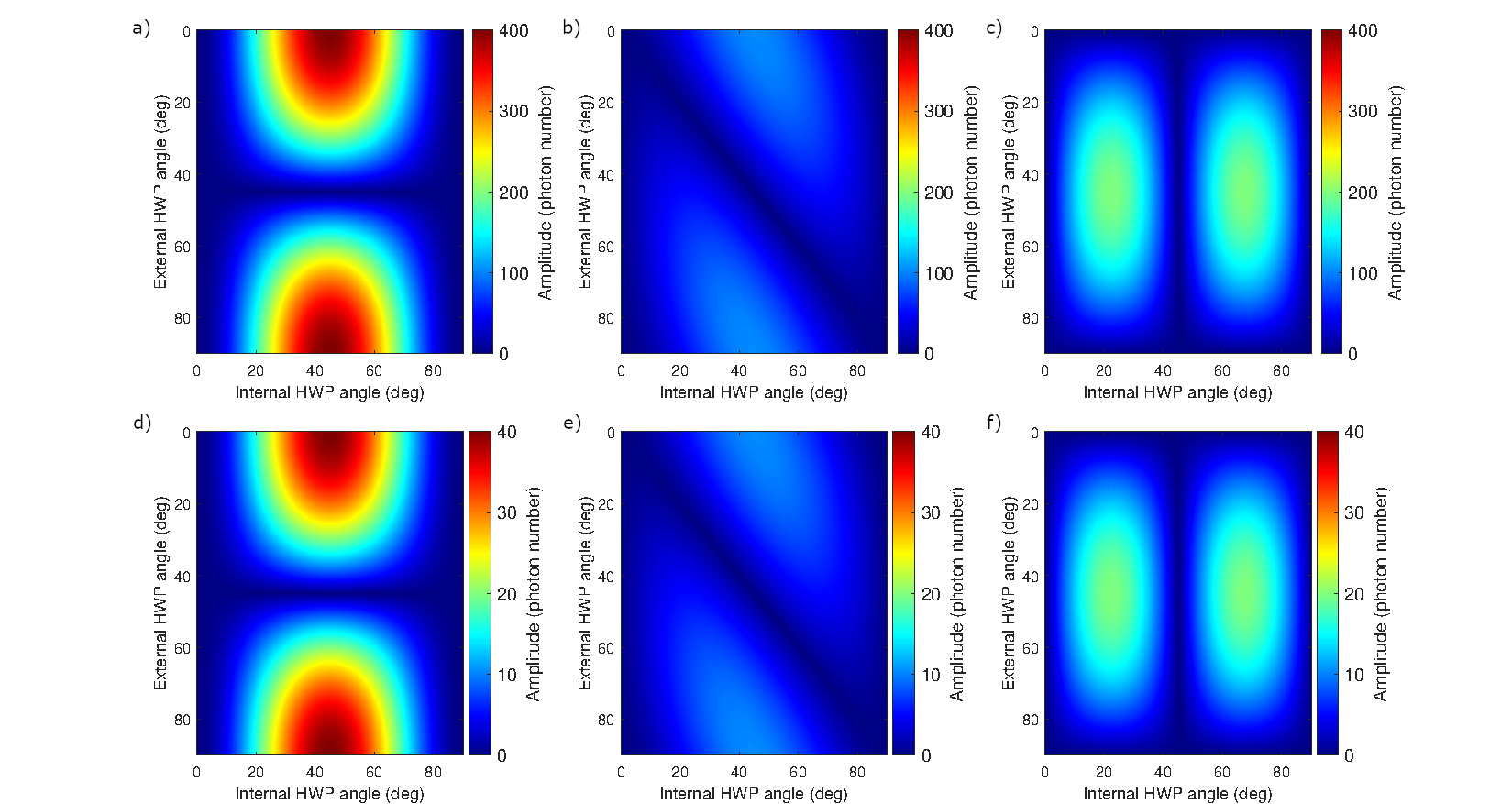}
\caption{Density plots of the interference fringe amplitude as a function of both of the HWP angles within the setup for three different idler mirror positions and two different gain balances. Upper plots show the amplitudes for (a) nonlinear interference (NI-IUP), (b) the mixed case, and (c) induced coherence (IC-IUP) respectively for $\xi_A == \xi_B$. Lower plots show the amplitudes for (d) NI-IUP, (e) the mixed case, and (f) IC-IUP respectively for $\xi_A = 0.1\xi_B$}
\label{fig:Balancing2}
\end{figure}

Figure~\ref{fig:Balancing1} shows colour density plots of the visibility as a function of both of the HWP angles within the setup. The top row (a-c) shows how the visibility varies at three different positions of the idler mirror, showing (a) NI-IUP, (b) a mixed case, and (c) IC-IUP. For a large unbalancing, for example $\xi_A = 0.1\xi_B$, shown in Fig.~\ref{fig:Balancing1}(d-f), it is still possible to regain full visibility in the case of the induced coherence setup by readjusting the balance of probabilities of signal photons from the two processes. However, this maximising of visibility comes at the expense of the reduced amplitude of oscillations, as shown in Fig.~\ref{fig:Balancing2}. This could be compensated for with the use of a second detector on the other exit port of the PBS (another advantage of the IC-IUP setup), which could improve the overall signal-to-noise.

\section{Interference mixing phase}

\begin{figure}[h!]
\centering\includegraphics[width = 15cm]{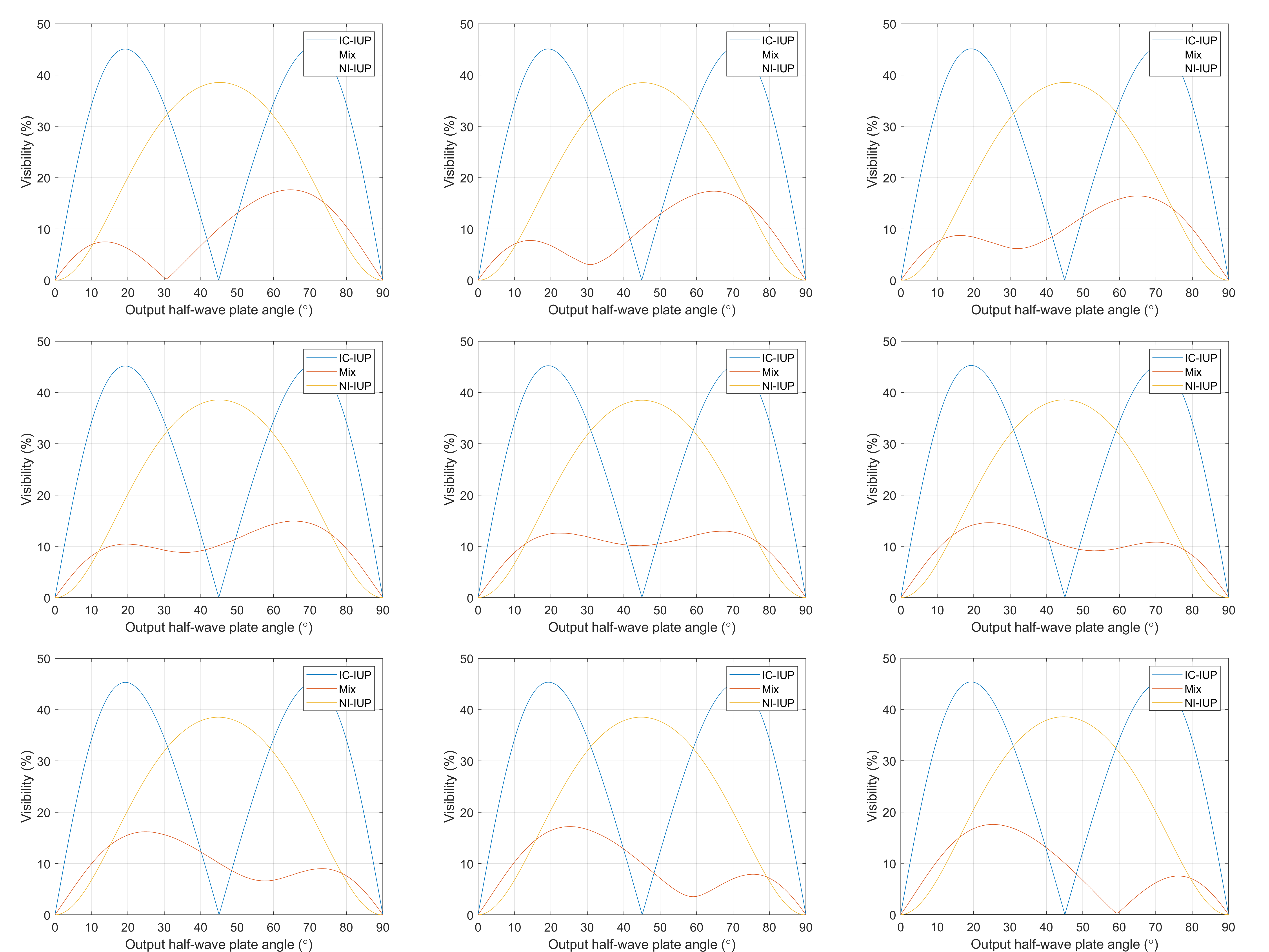}
\caption{Plots showing the how the visibility of the mixed case varies with external HWP angle at a range of additional path length introduced on the vertical signal mode (increasing from left to right and top to bottom, spanning a total path length shift of 362.5\,nm}
\label{fig:mixed_phase}
\end{figure}

As mentioned in the main text, by adding an additional path length to only the vertical component of the signal path (achieved experimentally by tilting the BBO crystal), one can introduce a phase shift between the two interference types. This can be seen clearly in the mixed case as shown in Figure~\ref{fig:mixed_phase} for a range of different added path lengths.

\section{Aliasing effects}

\begin{figure}[h!]
\centering\includegraphics[width = 12cm]{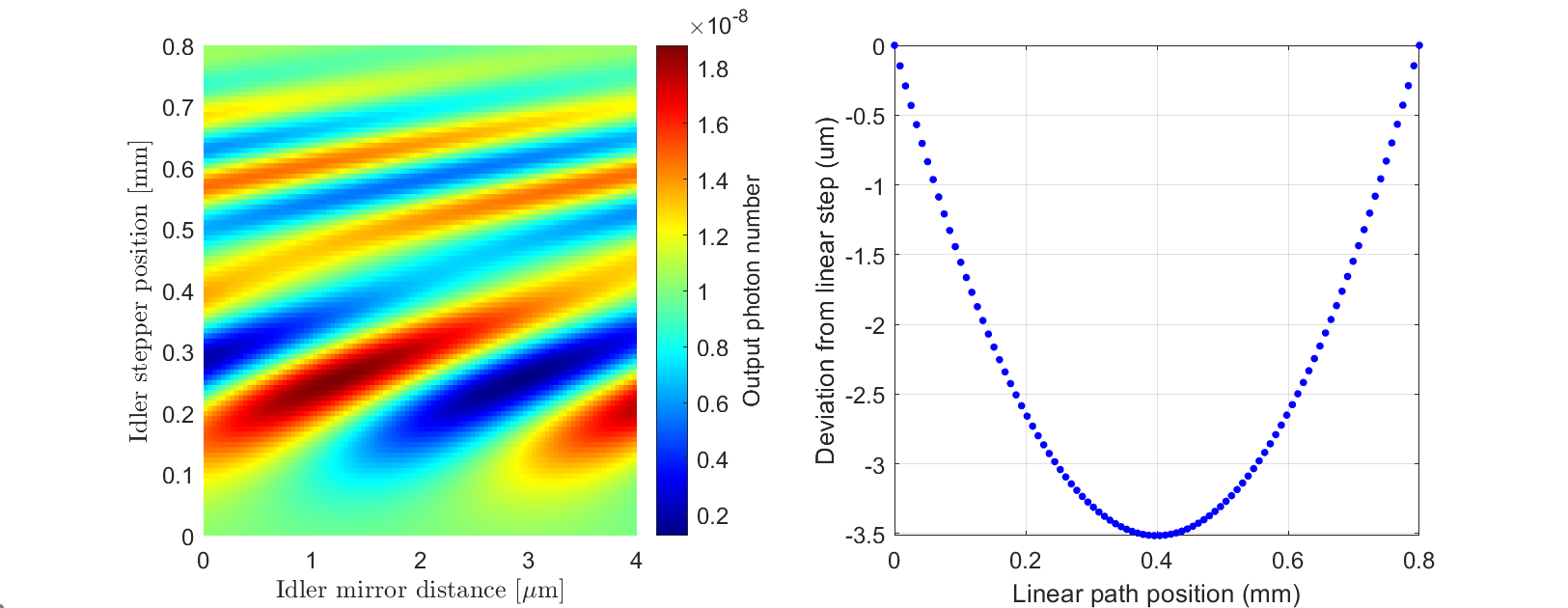}
\caption{Left: density plot showing theoretical signal count rates as a function of the large (nonlinear) changes in idler path length on the y-axis and fine shifts in idler path length on the x-axis. Right: plot of the deviation from linear motion in the step sizes on the y-axis of the left plot.}
\label{fig:Aliasing}
\end{figure}

While the phase in the experimental set up described here can be scanned with a closed loop piezo motor, scans over larger distances are performed with a stepper motor. This sampling of the interference fringes with larger step sizes makes aliasing effects apparent in the data. As seen in Fig.~3 in the main text, the fringe spacing on the y-axis in the experimental data appears to vary. This is not due to a difference in the frequency of the interference oscillation itself (as is evident from the period witnessed on the x axis). Instead, it is caused by inconsistencies in the step size of the stepper motor. On this axis of the data we are sampling the fringes at points spread apart much greater than the period of oscillation ($\sim10\,\mu$m). The oscillation we see on the y axis is therefore an aliased sampling. The stepper motor itself has a quoted unidirectional repeatability of $3.6\,\mu$m. As shown in Fig.~\ref{fig:Aliasing}, a small variation from linearity added to the stepper motion in the theoretical model consistent with this number produces an aliasing effect very similar to that of the experimental data.

\clearpage

\end{document}